%% file: MAPFFNNLO.tex
\documentclass[11pt,a4paper]{article}
\usepackage[colorlinks=true, linkcolor=black!50!blue, urlcolor=blue, citecolor=blue, anchorcolor=blue]{hyperref}
\usepackage[font=small,labelfont=bf,margin=0mm,labelsep=period,tableposition=top]{caption}
\usepackage[a4paper,top=2cm,bottom=2.5cm,left=2cm,right=2cm,bindingoffset=0mm]{geometry}

\usepackage{graphicx}
\usepackage{float}
\usepackage{afterpage}
\usepackage{epsfig,cite}
\usepackage{amssymb}
\usepackage{amsmath}
\usepackage{dsfont}
\usepackage{multirow}
\usepackage{url,hyperref}
\usepackage{booktabs}
\usepackage{tabularx}
\usepackage{xcolor}
\usepackage{bm}
\usepackage[symbol]{footmisc}
\usepackage{textcomp}
\usepackage{caption}
\usepackage[normalem]{ulem}

\begin{document}

\vspace{-2.0cm}
\begin{flushright}
JLAB-THY-22-3605
\end{flushright}
\vspace{.1cm}

\begin{center} 

  {\Large \bf Pion and kaon fragmentation functions\\
    \vspace{0.2cm}
    at next-to-next-to-leading order}
  \vspace{.7cm}

  Rabah Abdul Khalek$^1$, Valerio~Bertone$^2$, Alice Khoudli$^2$,
  Emanuele R. Nocera$^3$

\vspace{.3cm}
{\it ~$^1$ Jefferson Lab, Newport News, Virginia 23606, USA}\\
{\it ~$^2$ IRFU, CEA, Universit\'e Paris-Saclay, F-91191
  Gif-sur-Yvette, France}\\
{\it ~$^3$ The Higgs Centre for Theoretical Physics, University of Edinburgh,\\
JCMB, KB, Mayfield Rd, Edinburgh EH9 3JZ, Scotland\\}

\end{center}

\vspace{0.1cm}

\begin{center}
  {\bf \large Abstract}\\
\end{center}
  
We present a new determination of unpolarised charged pion and kaon
fragmentation functions from a set of single-inclusive electron-positron
annihilation and lepton-nucleon semi-inclusive deep-inelastic scattering data.
The determination includes next-to-next-to-leading order QCD corrections to
both processes, and is carried out in a framework that combines a neural-network
parametrisation of fragmentation functions with a Monte Carlo representation of
their uncertainties. We discuss the quality of the determination, in particular
its dependence on higher order corrections. 

\vspace{1cm}

\section{Introduction}\label{sec:intro}

In a recent paper~\cite{Khalek:2021gxf}, we presented a determination
of the fragmentation functions (FFs)~\cite{Metz:2016swz} of charged pions from
an analysis of hadron-production measurements in single-inclusive
electron-positron annihilation (SIA) and semi-inclusive deep-inelastic
scattering (SIDIS). The analysis, accurate to next-to-leading order
(NLO) in perturbative quantum chromodynamics (QCD), utilised a framework
that combines a neural network parametrisation of FFs (optimised through
knowledge of the analytical derivative of neural networks with respect to their
parameters) with a Monte Carlo representation of FF uncertainties.
This approach --- which has been extensively used by the NNPDF Collaboration
to determine the parton distribution functions (PDFs) of the
proton~\cite{Nocera:2014gqa,NNPDF:2014otw,NNPDF:2017mvq,Ball:2021leu} and of
nuclei~\cite{AbdulKhalek:2019mzd,AbdulKhalek:2020yuc,Khalek:2022zqe} ---
allowed us to reduce model bias in FF parametrisation as much as possible,
and to faithfully propagate experimental and PDF uncertainties into FFs.
These features are essential to achieve the methodological accuracy of FFs
that are utilised to analyse, {\it e.g.}, high-precision hadron
production measurements at the Large Hadron Collider (LHC) and, in the
future, at the Electron Ion Collider (EIC)~\cite{AbdulKhalek:2021gbh}.

Methodological accuracy is however only one component of the overall accuracy
of the FF determination. Other important components are the accuracy of the
experimental and theoretical inputs that also enter the FF determination.
Regarding experimental accuracy, interplay between SIA and SIDIS measurements
was studied at length in Ref.~\cite{Khalek:2021gxf}, and the latter were found
to be essential to constrain FFs for individual quark flavours. The two classes
of measurements are indeed sensitive to different quark FF combinations due to
the way in which the corresponding observables factorise~\cite{Collins:1989gx}.
Regarding theoretical accuracy, derivation of approximate
next-to-next-to-leading order (NNLO) corrections to SIDIS, obtained from
expansion of the resummed expressions, have been completed a few months
ago~\cite{Abele:2021nyo}.\footnote{Approximate N$^3$LO corrections have also
  been presented in Ref.~\cite{Abele:2022wuy}.} Given the long-standing
availability of NNLO corrections to
SIA~\cite{Rijken:1996vr,Rijken:1996npa,Rijken:1996ns},
and to time-like evolution~\cite{Mitov:2006ic,Moch:2007tx,Almasy:2011eq}, it is
therefore natural to extend the framework developed in
Ref.~\cite{Khalek:2021gxf} to NNLO. This is the goal of this paper, in which we
complement the original pion {\sc MAPFF1.0} FF sets~\cite{Khalek:2021gxf} with
their NNLO counterparts.

We also produce analogous kaon FF sets, both at NLO and NNLO. Together with
pions, kaons represent the most copiously produced hadrons in high-energy
particle collisions. An accurate knowledge of kaon FFs is of crucial importance
to use SIDIS measurements (including when the initial-state proton is
longitudinally polarised) to constrain the (polarised) strange quark and
anti-quark PDFs.

The {\sc MAPFF1.0} pion and kaon FF sets presented in this paper extend the
available NNLO analyses that are based solely on SIA
measurements~\cite{Anderle:2016czy,Bertone:2017tyb}, as well as the very recent
(and to date only) NNLO global analysis of pion FFs based on SIA and SIDIS
measurements~\cite{Borsa:2022vvp}. As for the previous
determination~\cite{Khalek:2021gxf}, the NLO and NNLO {\sc MAPFF1.0} pion and
kaon FF sets are publicly delivered through the {\sc LHAPDF}
library~\cite{Buckley:2014ana}. The software developed to produce them is also
made open source~\cite{rabah_abdul_khalek_2022_6264693}.
In Sect.~\ref{sec:input} we summarise the experimental, theoretical and
methodological input to our analysis; in Sect.~\ref{sec:results} we discuss
the main results; and in Sect.\ref{sec:conclusions} we present a summary and an
outlook.

\section{Experimental, theoretical, and methodological input}
\label{sec:input}

The SIA and SIDIS experimental measurements that are used as input to this
analysis closely follow those of our previous work. For pions, we use
exactly the same measurements as in Ref.~\cite{Khalek:2021gxf}, albeit with a
different treatment of experimental uncertainties for the COMPASS
data~\cite{COMPASS:2016xvm}, see below. For kaons, we use SIA measurements
performed at CERN (ALEPH~\cite{ALEPH:1994cbg}, DELPHI~\cite{DELPHI:1998cgx} and
OPAL~\cite{OPAL:1994zan}), DESY
(TASSO~\cite{TASSO:1980dyh,TASSO:1982bkc,TASSO:1988jma}),
KEK (BELLE~\cite{Belle:2013lfg} and TOPAZ~\cite{TOPAZ:1994voc})
and SLAC (BABAR~\cite{BaBar:2013yrg}, TPC~\cite{TPCTwoGamma:1988yjh}
and SLD~\cite{SLD:2003ogn}); we also use SIDIS measurements performed at CERN
by COMPASS~\cite{COMPASS:2016crr} and at DESY by HERMES~\cite{HERMES:2012uyd}.

In the case of SIA, the data corresponds to the sum of the cross
section for the production of positively and negatively charged kaons,
differential with respect to either the longitudinal momentum fraction $z$
of the outgoing kaon carried by the fragmenting parton or the momentum of the
measured kaon (see Sect.~2.2 in Ref.~\cite{Bertone:2017tyb} for details).
For BELLE, we use the measurement corresponding to an integrated luminosity
$\mathcal{L}=68$~fb$^{-1}$~\cite{Belle:2013lfg}. A more recent measurement,
based on a larger luminosity $\mathcal{L}=558$~fb$^{-1}$,
exists~\cite{Belle:2020pvy}. However we do not consider it because of a poor
control of the degree of correlation of systematic uncertainties, which
typically exceed in magnitude uncorrelated statistical uncertainties
(see Ref.~\cite{Khalek:2021gxf}). For BABAR we use the {\it conventional}
data set, as done in other analyses, see
{\it e.g.} Refs.~\cite{deFlorian:2017lwf,Bertone:2017tyb,Moffat:2021dji}.
This is in contrast to the pion measurement, for which we use the {\it prompt}
data set. The difference between the prompt and conventional data sets is that
only primary hadrons or decay products from particles with lifetime $\tau$
shorter than about 10$^{-11}$~s are retained in the former. While prompt and
conventional cross sections differ by about 5--15\% for pions, they are
almost identical for kaons. As shown in previous analyses of pion
FFs~\cite{deFlorian:2014xna,Sato:2016wqj,Bertone:2017tyb}, the inclusion of
the conventional pion data set leads to a significant deterioration of the
statistical quality of the fit. For this reason we do not consider the
BABAR conventional pion data set at all. For DELPHI and SLD, in addition to the
inclusive measurements, we also consider flavour-tagged measurements, whereby
the production of the observed kaon has been reconstructed from hadronisation of
all light quarks or of a $b$ quark. 

In the case of SIDIS, the data corresponds to the hadron multiplicity,
{\it i.e.} the SIDIS cross section normalised to the corresponding inclusive
DIS cross section (see Sect.~3 in Ref.~\cite{Khalek:2021gxf} for details).
For HERMES, similarly to what we did in the case of pions, we consider the
projection of the fully differential measurement as a function of the
transferred energy $Q^2$ and of $z$ in individual bins of the momentum fraction
$x$ carried by the incoming parton. We discard the bins with $z<0.2$, which are
used to control the model dependence of the smearing-unfolding procedure, and
with $z>0.8$, which lie in the region where the fractional contribution from
exclusive processes is sizeable.

The kinematic coverage of the pion and kaon data is similar, see Sect.~2
of Ref.~\cite{Khalek:2021gxf} for a detailed discussion. Kinematic cuts, to
select only data points for which perturbative fixed-order predictions
are reliable, are as in Ref.~\cite{Khalek:2021gxf} for pions. Specifically,
for SIA we retain only the data points that fall in the interval
$[z_{\rm min}, z_{\rm max}]$, with $z_{\rm min}=0.02$ for experiments at a
centre-of-mass energy equal to $M_Z$ and $z_{\rm min}=0.075$ for all other
experiments, and $z_{\rm max}=0.9$ for all experiments. For SIDIS, we retain only
the data points satisfying $Q>Q_{\rm cut}$, with $Q_{\rm cut}=2$~GeV. In the case
of kaons, we adopt exactly the same kinematic cuts as in the case of pions, with
the exception of the value of $z_{\min}$ used for the BELLE and BABAR
experiments, which is set to 0.2. The reason being that the onset of small-$z$
corrections at the centre-of-mass energy of $B$ factories occurs for kaons at a
higher value of $z$ than it does for pions. We use the same set of cuts in the
NLO and NNLO fits. In principle, different cuts could be defined depending on
the perturbative order to maximise the amount of experimental information
included in the fit. However, we prefer to be conservative, and use the same
cuts determined in Ref.~\cite{Khalek:2021gxf}, where in particular a scan of the
fit quality upon variation of $Q_{\rm cut}$ was performed. A similar study will
be discussed further below, after which we will show that slightly less
restrictive SIDIS cuts could be used without significantly spoiling the fit
quality. However, these may alter the FF accuracy or precision, as we will also
discuss.

Information on correlations of experimental uncertainties is taken into account
whenever available, as detailed in Sect.~2 of Ref.~\cite{Khalek:2021gxf}. In
contrast to Ref.~\cite{Khalek:2021gxf}, however, we no longer consider the
systematic uncertainty for the COMPASS
measurements~\cite{COMPASS:2016xvm,COMPASS:2016crr}
to be 100\% correlated across bins. We instead implement the recommendation
to split the systematic uncertainty into two components, and take only the
largest component (which amounts to 80\% of the total systematic uncertainty) to
be 100\% correlated across bins. The remaining component is treated as fully
uncorrelated, and is added in quadrature to the statistical uncertainty. This
treatment is applied equally to pion and kaon measurements. In this respect,
the NLO fit of pion FFs that we will present below differs from that
of Ref.~\cite{Khalek:2021gxf}. Further below we will also discuss how the fit
quality and FFs are affected by variations in the treatment of experimental
correlations in the COMPASS measurements.

The theoretical setup of our analysis closely follows that discussed in Sect.~3
of Ref.~\cite{Khalek:2021gxf}. New to this determination is the inclusion of
NNLO corrections to time-like DGLAP evolution equations and to SIA and SIDIS
coefficient functions. Corrections to evolution equations and to SIA build upon
various implementations and benchmarks carried out in previous
work~\cite{Bertone:2017tyb,Bertone:2015cwa,Bertone:2013vaa,Bertone:2017gds}.
Corrections to SIDIS are instead taken from Ref.~\cite{Abele:2021nyo}. These
corrections were derived using the threshold-resummation formalism.
They are therefore approximate in that they only include the dominant
contributions associated with the emission of soft
gluons. Nevertheless, they are sufficiently accurate for our purpose:
in Ref.~\cite{Abele:2021nyo} it was shown that the approximate and
exact NLO results, which are both known, differ only by a tiny amount
in the kinematic region relevant to the SIDIS data considered in this
work (see Fig.~1 in Ref.~\cite{Abele:2021nyo}). The size of this
difference is always largely smaller than the size of experimental
uncertainties. It is reasonable to argue that the approximate NNLO
result would not differ from the exact NNLO result, if known, by a
larger amount. The unquantified uncertainty due to utilising the
approximate, in lieu of the exact, result is therefore practically
negligible.  Furthermore, the inclusion of the dominant NNLO terms is
expected to reduce the dependence of the cross section on the
renormalisation and factorisation scales. This is what happens, as
also shown in Ref.~\cite{Abele:2021nyo}, see in particular Fig.~3
there. Finally, excellent perturbative stability of the SIDIS cross
section was found very recently by extending the derivation in
Ref.~\cite{Abele:2021nyo} to N$^3$LO~\cite{Abele:2022wuy}.

As in Ref.~\cite{Khalek:2021gxf}, we use the
NNPDF3.1~\cite{NNPDF:2017mvq} PDF sets as input to the computation of
SIDIS cross sections, specifically those obtained assuming that charm
is perturbatively generated. The perturbative order of the PDF set is
taken consistently with the perturbative order of the FF analysis. We
have explicitly verified, {\it e.g.} by using the more recent NNPDF4.0
parton sets~\cite{Ball:2021leu}, that the dependence of our results on
the choice of the PDFs is very weak, due to cancellations that occur
in the multiplicity ratio, see also
Ref.~\cite{Khalek:2021gxf}. Because no heavy-quark mass corrections
have been determined for SIDIS, our analysis is carried out in the
zero-mass variable-flavour-number scheme. In this scheme all active
partons are treated as massless, but a partial heavy-quark mass
dependence is introduced by requiring that sub-schemes with different
numbers of active flavours match at the heavy-quark thresholds. The
values of the charm- and bottom-quark thresholds are set to
$m_c=1.51$~GeV and $m_b=4.92$~GeV, respectively, consistently with the
NNPDF3.1 input PDF sets. Heavy-quark FFs are not set to zero below
their respective thresholds, but are mathched\footnote{In both the NLO
  and the NNLO fits, FF heavy-quark matching conditions are
  implemented to $\mathcal{O}(\alpha_s)$, \textit{i.e.} NLO, using the
  results of Ref.~\cite{Cacciari:2005ry}. Indeed, the
  $\mathcal{O}(\alpha_s^2)$, \textit{i.e.} NNLO, corrections to the
  matching conditions are currently unknown.} and kept constant, {\it
  i.e.} they do not evolve. Their contribution is suppressed by PDFs
in SIDIS (see Sect.~3 in Ref.~\cite{Khalek:2021gxf} for details). We
finally note that isoscalarity of the SIDIS targets is taken into
account by assuming exact SU(2) symmetry between protons and neutrons;
no nuclear corrections are taken into account, as are no target or
hadron mass corrections.

The statistical framework used in this analysis to infer FFs from experimental
data is also the same as in Ref.~\cite{Khalek:2021gxf}. Ingredients of the
framework are the representation of experimental uncertainties into FFs by
means of Monte Carlo sampling, and the parametrisation of FFs by means of neural
networks. In the first respect, all of our FF sets are made of $N_{\rm rep}=200$
Monte Carlo replicas. In the case of SIDIS, a different PDF replica is chosen
at random from the NNPDF3.1 parton set for each fitted FF replica. This ensures
the propagation of PDF uncertainties into FFs. In the second respect, we
consider, separately for pions and kaons, a single one-layered feed-forward
neural network with one input node corresponding to the momentum fraction $z$,
20 intermediate nodes with a sigmoid activation function, and 7 output nodes
with a linear activation function. This architecture amounts to a
total of 187 parameters.

The output nodes correspond to the independent FFs of the positively charged
hadrons that we fit. In the case of pions, these are given by Eq.~(10)
in Ref.~\cite{Khalek:2021gxf}. In the case of kaons, these are obtained from
those for pions by exchanging $d$ and $s$ quarks, that is:
\begin{equation}
  \{D^{K^+}_u, \quad
  D^{K^+}_{\bar{s}}, \quad
  D^{K^+}_{s}=D^{K^+}_{\bar{u}}, \quad
  D^{K^+}_d=D^{K^+}_{\bar{d}}, \quad
  D^{K^+}_c=D^{K^+}_{\bar{c}}, \quad
  D^{K^+}_b=D^{K^+}_{\bar{b}}, \quad
  D^{K^+}_g \}\,;
  \label{eq:param_flavours}
\end{equation}
FFs for negatively charged hadrons are obtained from the positively charged ones
by charge conjugation. The output nodes are squared to avoid large,
unphysically negative FFs. The parametrisation is introduced at the initial
scale $\mu_0=5$~GeV, as in our previous analysis~\cite{Khalek:2021gxf}. Because
this value is above the bottom quark threshold, we can parametrise the charm-
and bottom-quark FFs, which receive large non-perturbative contributions, on the
same footing as light quark FFs; FFs are then evolved to the scale of the SIDIS data
(which can be lower than $Q_0$) in our fit. Our parametrisation does not include
any power-like function to control the low- and high-$z$ behaviours of
FFs; however we require them to vanish at $z=1$ by subtraction of the neural network itself,
see also Ref.~\cite{Bertone:2017tyb}.

Optimisation of the neural network parameters is achieved by minimisation of the
$\chi^2$, see {\it e.g.} Eq.~(21) in Ref.~\cite{Khalek:2021gxf} for the exact
definition used. Cross-validation is used to avoid overfitting, with a 50\%
training fraction for all of the data sets that contain more than 10 data 
points, otherwise the training fraction is 100\%. Minimisation is realised
with the Levenberg-Marquardt algorithm as implemented in the
{\sc Ceres Solver} package~\cite{ceres-solver}; analytical
derivatives with respect to the parameters of the neural network are
provided by the {\sc NNAD} library~\cite{AbdulKhalek:2020uza}.

\section{Results and discussion}
\label{sec:results}

In Table~\ref{tab:chi2} we report the number of data points, $N_{\rm dat}$, and
the value of the $\chi^2$ per data point, $\chi^2/N_{\rm dat}$, for each data set
included in our pion and kaon fits at NLO and NNLO. Values
corresponding to the SIA, SIDIS, and global data sets are also displayed.
Inspection of Table~\ref{tab:chi2} allows us to draw two observations.

\begin{table}[!t]
  \scriptsize
  \centering
  \renewcommand{\arraystretch}{1.4}
  \input{_tables/chi2.tex}\\
  \vspace{0.5cm}
  \caption{The number of data points, $N_{\rm dat}$, and the $\chi^2$ per data
    point, $\chi^2/N_{\rm dat}$, for each hadronic species and perturbative order
    considered in the fits of this analysis.}
  \label{tab:chi2}
\end{table}

First, we notice that the fit quality, as measured by the $\chi^2$ per data
point, reveals a generally good description of the entire data set, for both
pions and kaons, and separately for SIA and SIDIS measurements. Anomalously
small values of the $\chi^2$ per data point are found for some data sets, that
have low statistical significance because of either their limited number of data
points (TASSO and HERMES) or their large uncorrelated uncertainties (BELLE),
see Refs.~\cite{Khalek:2021gxf,Bertone:2017tyb}. The fit quality of the NLO
pion fit is better than that found in Ref.~\cite{Khalek:2021gxf}. The two
fits, albeit based on the same data set and methodology, differ for the
treatment of correlations in the COMPASS measurement. This is the reason for
the reduction of the total $\chi^2$ per data point from 0.90
in Ref.~\cite{Khalek:2021gxf} to 0.68, as we will further discuss below.

Second, we notice that the dependence of the fit quality on the perturbative
accuracy of the fit is opposite for pion and kaon FFs. When moving from NLO to
NNLO, the total $\chi^2$ per data point deteriorates from 0.68 to 0.76 in the
former case, while it improves from 0.62 to 0.55 in the latter case. This
behaviour is somewhat surprising given that NNLO corrections are independent
from the hadron species. Also, the deterioration equally affects SIA and SIDIS
data for pions, as does the improvement for kaons. The reasons for this
behaviour, which differs from what was observed in a similar NNLO
analysis~\cite{Borsa:2022vvp}, are only partly understood, as we will discuss
below.

\begin{figure}[!t]
  \centering
  \includegraphics[width=\textwidth]{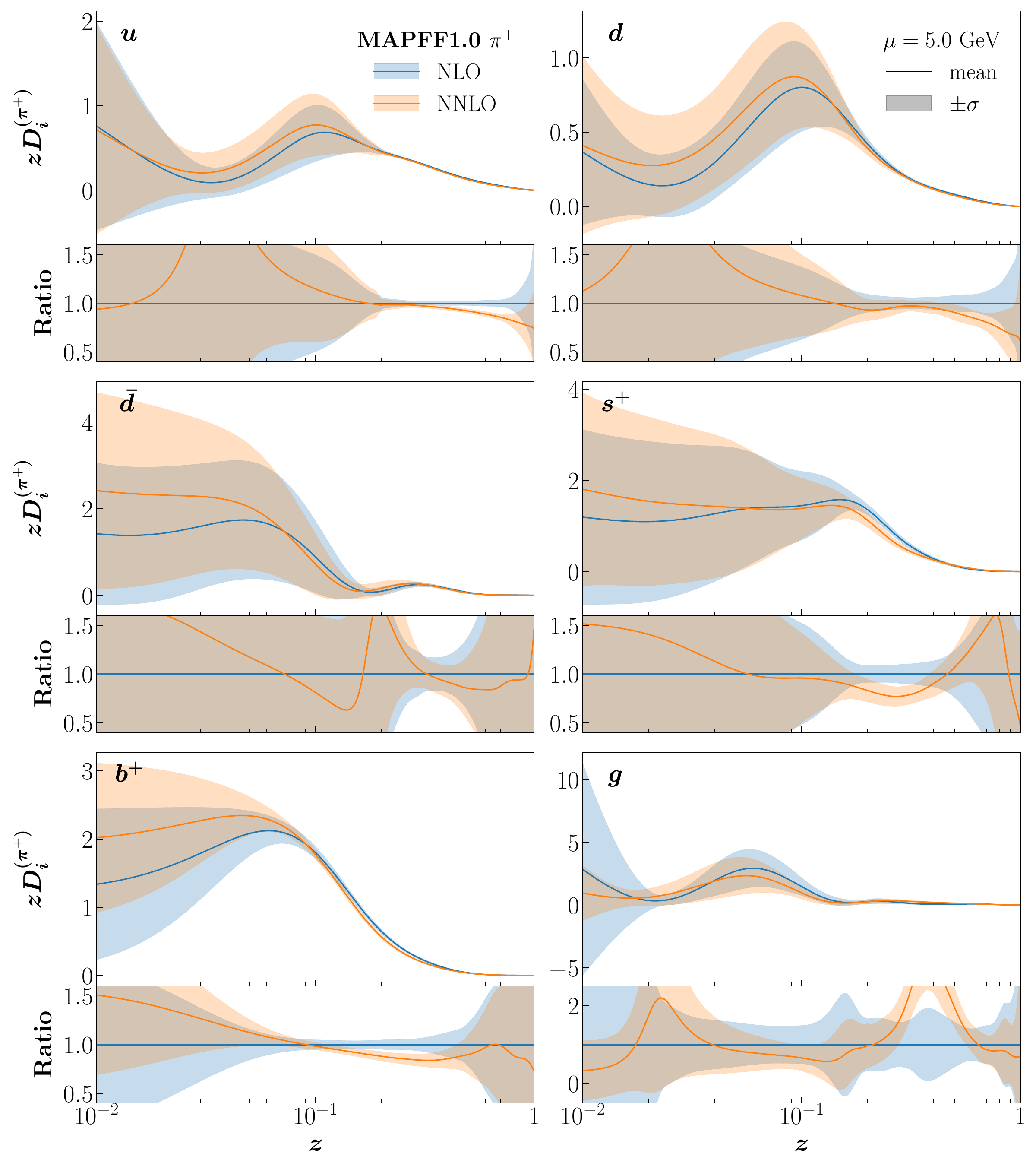}\\
  \vspace{0.5cm}
  \caption{Comparison of the NLO and NNLO FFs for positively charged pions. We
    display the $D_u^{\pi^+}$, $D_d^{\pi^+}$, $D_{\bar d}^{\pi^+}$, $D_{s^+}^{\pi^+}$,
    $D_{b^+}^{\pi^+}$ and $D_{g}^{\pi^+}$ FFs at $\mu=5$~GeV. Expectation values and
    uncertainties correspond to the mean and standard deviation computed over
    the ensemble of FF replicas. For each FF we plot the absolute distributions
    in the upper panels and their ratio to the central value of the NLO FFs in
    the lower ones.}
  \label{fig:pions_NLOvsNNLO}
\end{figure}

\begin{figure}[!t]
  \centering
  \includegraphics[width=\textwidth]{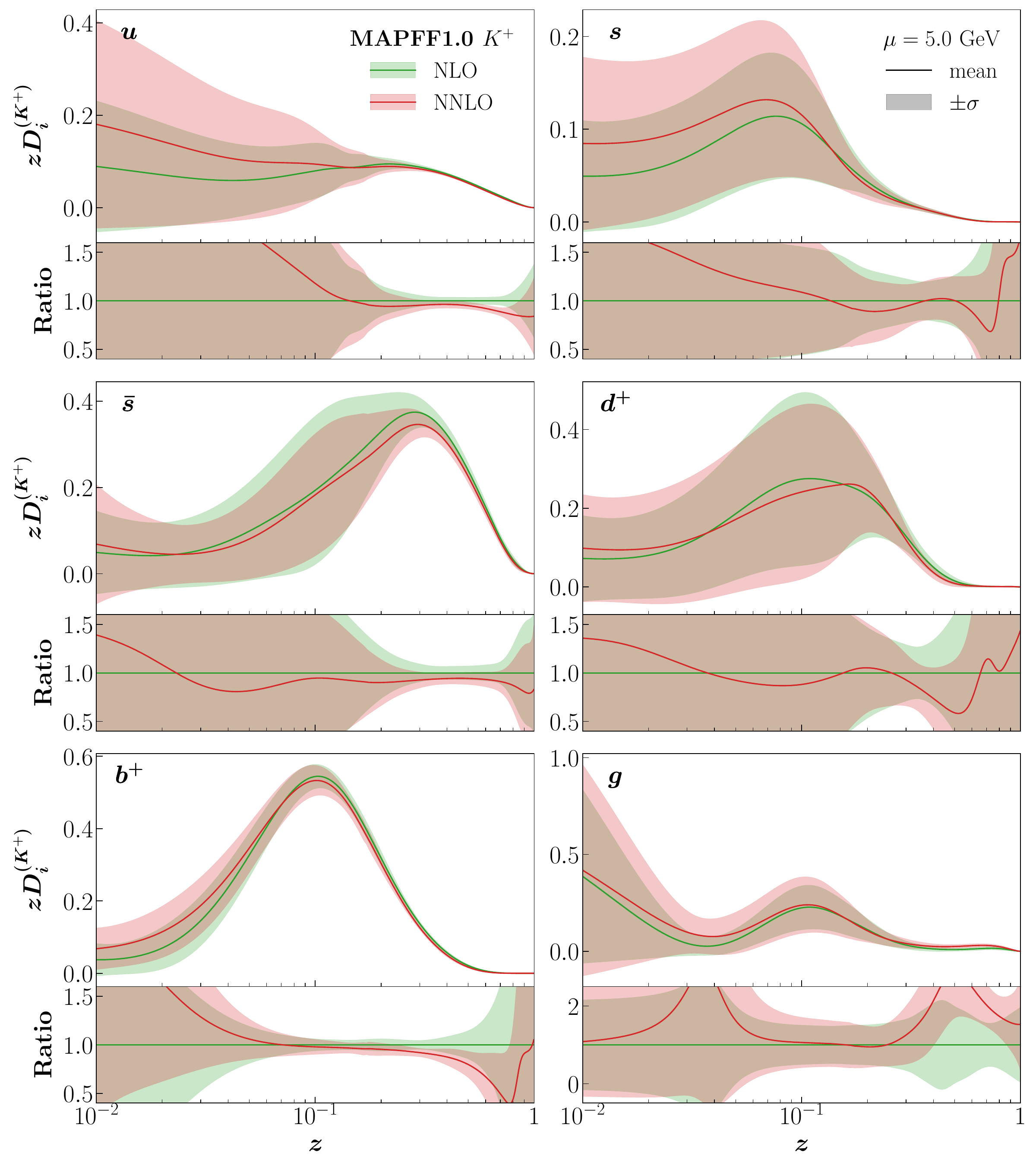}\\
  \vspace{0.5cm}
  \caption{Same as Fig.~\ref{fig:pions_NLOvsNNLO} for kaons, now displaying
    $D_u^{K^+}$, $D_s^{K^+}$, $D_{\bar s}^{K^+}$, $D_{d^+}^{K^+}$, $D_{b^+}^{K^+}$ and
    $D_{g}^{K^+}$.}
  \label{fig:kaons_NLOvsNNLO}
\end{figure}

Figures~\ref{fig:pions_NLOvsNNLO} and \ref{fig:kaons_NLOvsNNLO} display,
for positively charged pions and kaons respectively, a comparison of the FFs
(times the longitudinal momentum fraction $z$) obtained from our NLO and NNLO
fits. For pions, we show $D_u^{\pi^+}$,
$D_d^{\pi^+}$, $D_{\bar d}^{\pi^+}$, $D_{s^+}^{\pi^+}$, $D_{b^+}^{\pi^+}$ and
$D_{g}^{\pi^+}$; for kaons, we show $D_u^{K^+}$,
$D_s^{K^+}$, $D_{\bar s}^{K^+}$, $D_{d^+}^{K^+}$, $D_{b^+}^{K^+}$ and
$D_{g}^{K^+}$. In both cases, FFs are displayed at the parametrisation scale
$\mu=5$~GeV, their expectation values and uncertainty bands correspond to the
mean and standard deviation computed over the ensemble of FF replicas, and the
lower insets display the FFs normalised to the NLO FFs.

Inspection of Figs.~\ref{fig:pions_NLOvsNNLO} and~\ref{fig:kaons_NLOvsNNLO}
reveals that inclusion of NNLO corrections results in a suppression of quark
FFs, and in an enhancement of the gluon FF in the large-$z$ region,
$z\gtrsim 0.5$. This behaviour is expected: quark FFs ought to be suppressed to
counteract the enhancement of theoretical predictions for SIA and SIDIS cross
sections induced by NNLO corrections~\cite{Anderle:2015lqa,Abele:2021nyo}.
At the same time, the gluon FF is enhanced to accommodate stronger evolution
effects. The size of the suppression depends on the quark flavour and on the
hadron species: the instances in which this is more marked are for $D_u^{\pi^+}$,
$D_d^{\pi^+}$, $D_{b^+}^{\pi^+}$, $D_u^{K^+}$, and $D_{b^+}^{K^+}$. In these cases, the
suppression can be as large as 10-20\%. By comparison, the quark FF uncertainty
is only about a few percent. The enhancement of the gluon FF can be as large as
60\%. Because the uncertainties on the gluon FF are significantly larger than
for quark FFs, the enhancement is just such that the outer edges of the NLO and
NNLO uncertainty bands touch each other: this corresponds to a $\sqrt{2}$
difference of a standard deviation. Be that as it may, in all of these cases
the impact of NNLO corrections on FFs is statistically significant.

While the qualitative effect of NNLO corrections on the FFs displayed in
Figs.~\ref{fig:pions_NLOvsNNLO}-\ref{fig:kaons_NLOvsNNLO} is expected, their
quantitative effect on the pion FF fit is more difficult to interpret. In
particular, it is surprising that the inclusion of NNLO corrections does not
improve the fit quality, as already observed above. In an attempt to investigate
the reason(s) for this behaviour, we have carried out a set of additional
studies, which also served the purpose to test the stability of our results.

\begin{figure}[!t]
  \centering
  \includegraphics[width=0.49\textwidth]{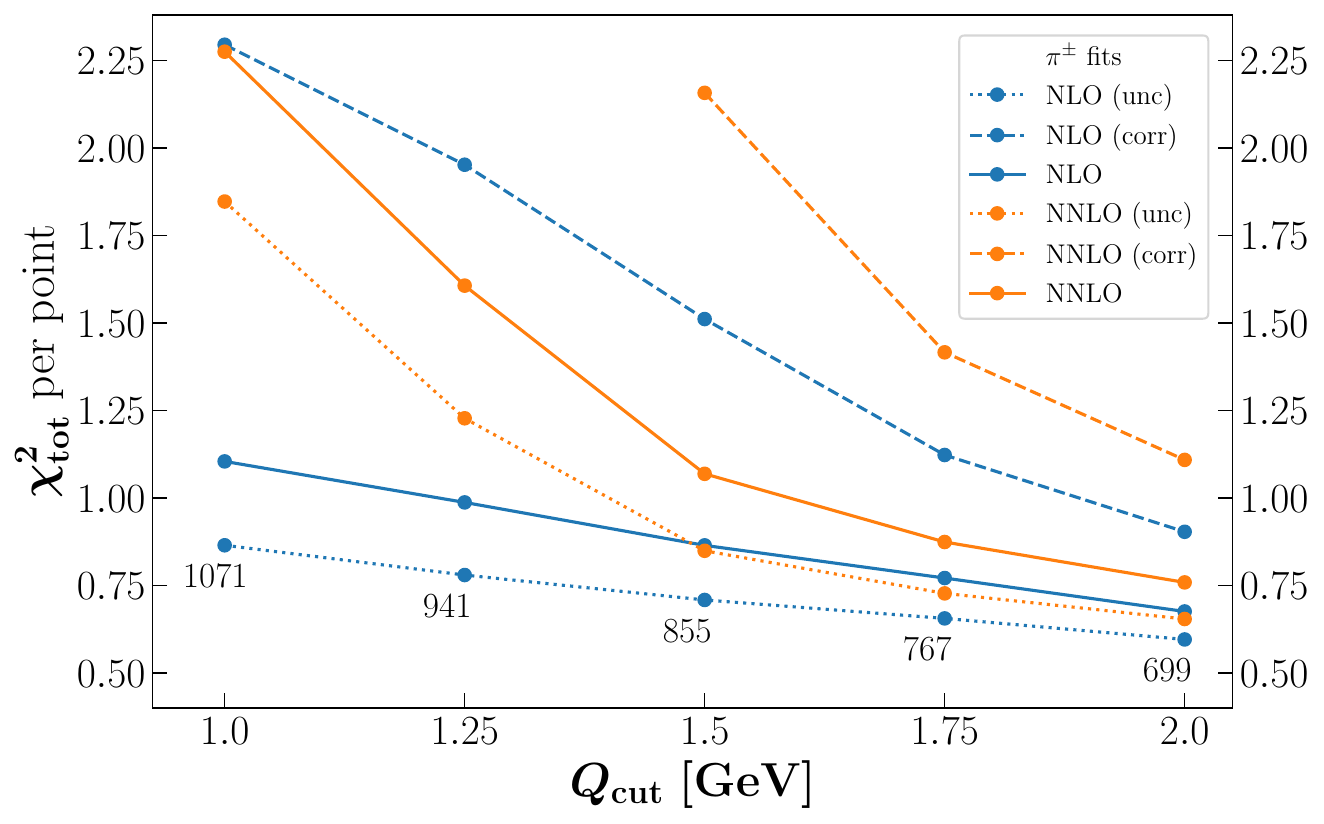}
  \includegraphics[width=0.49\textwidth]{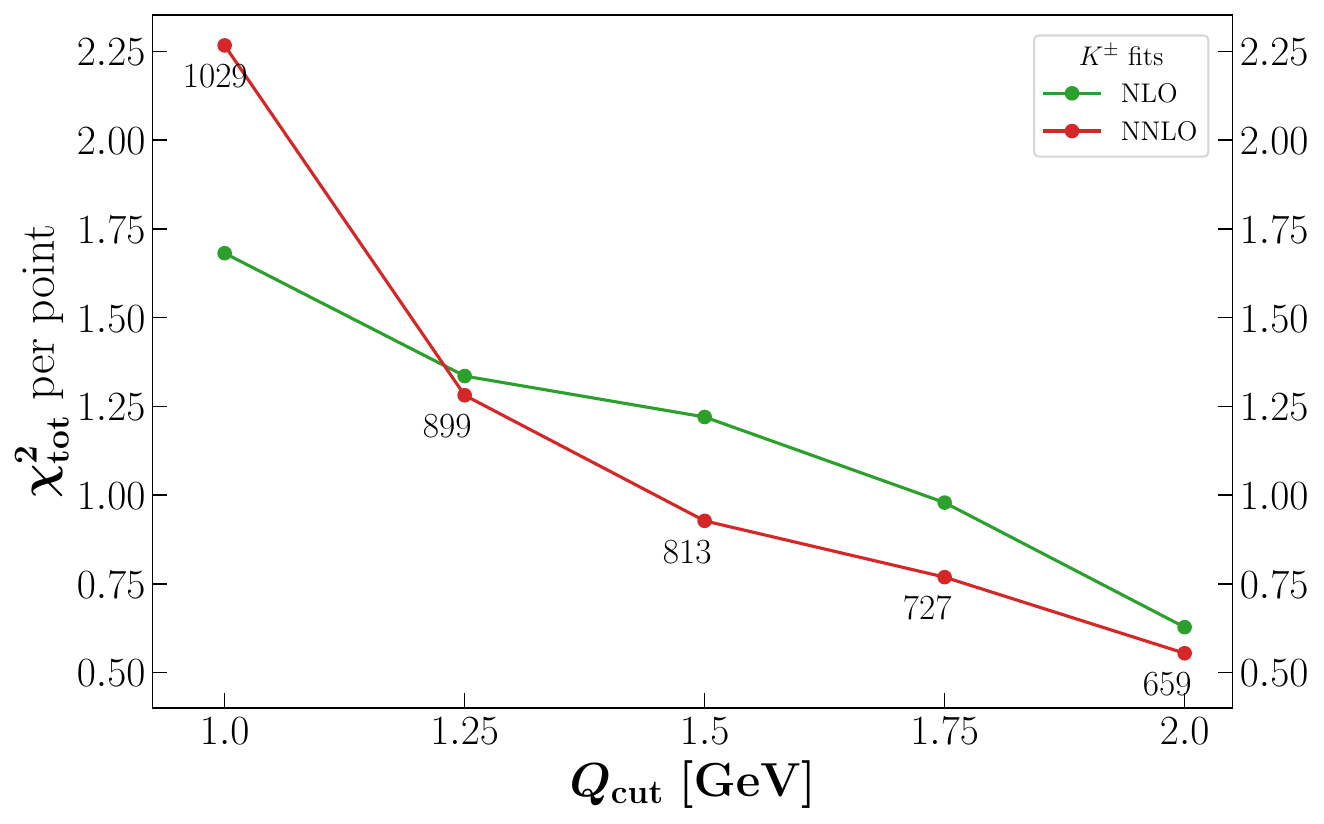}\\
  \vspace{0.5cm}
  \caption{The value of the total $\chi^2$ per data point as a function of the
    cut on $Q$, $Q_{\rm cut}$, applied to the SIDIS data in the pion (left) and
    kaon (right) FF fits. For each value of $Q_{\rm cut}$, the number of data
    points included in the fits are also displayed. Both NLO and NNLO fits are
    considered. In the case of the fit of pion FFs, various correlation models
    for the COMPASS data are taken into account, see the text for details.}
  \label{fig:chi2_scan}
\end{figure}

The first of such studies consists in varying the kinematic cut on the
virtuality $Q$, $Q_{\rm cut}$, in the analysis of SIDIS data. The study is
similar to the one carried out in Sect.~5.3.3 of Ref.~\cite{Khalek:2021gxf}:
we have repeated our NLO and NNLO fits, for both pions and kaons, varying the
value of $Q_{\rm cut}$ in the range [1.00, 2.00]~GeV in steps of 0.25~GeV. The
value of the $\chi^2$ per data point corresponding to the global data set for
each of these fits is displayed in Fig.~\ref{fig:chi2_scan} for pions (left)
and for kaons (right). The number of data points included in each fit is also
indicated.

As one can see from Fig.~\ref{fig:chi2_scan}, the value of the $\chi^2$ per
data point increases as the value of $Q_{\rm cut}$ is lowered, irrespective of
the hadron species. Interestingly, for both pions and kaons, the rise is
steeper at NNLO than at NLO; that is, the fit quality of the NNLO fit
deteriorates much faster than that of the NLO fit as the value of $Q_{\rm cut}$
is decreased. While for pions, as already noted, the global $\chi^2$ per data
point is always larger at NNLO than at NLO, for kaons the rise is such that
the quality of the NNLO fit becomes worse than that of the NLO fit if
$Q_{\rm cut}=1$~GeV.

We have nevertheless verified that, in our framework, the fit quality of the
pion FFs at NNLO is always better than its NLO counterpart if SIA or SIDIS data
sets are fitted separately. This at least confirms that the expected
perturbative convergence is recovered for each individual process. These fits
have however exposed how relevant the interplay between SIA and SIDIS
measurements is. On the one hand, the pion FFs determined by fitting only SIA
data cannot be used to predict SIDIS data, because FFs for different quark
flavours cannot be disentangled, see {\it e.g.} Sect.~3 in
Ref.~\cite{Bertone:2017tyb} and Sect.~3 in Ref.~\cite{Khalek:2021gxf}. On the
other hand, the pion FFs determined by fitting only SIDIS data do not provide a
good description of SIA data, because the two classes of measurements probe
somewhat disconnected kinematic regions, see {\it e.g.} Fig.~1 in
Ref.~\cite{Khalek:2021gxf}. In particular, SIDIS measurements
probe FFs at rather low energies (a few~GeV); they should therefore be evolved
at higher energies (up to the $Z$-boson mass) with large extrapolation
uncertainties. The reason why the quality of the pion FF NNLO fit becomes
consistently worse than its NLO counterpart when SIA and SIDIS measurements are
fitted together, be it the inconsistency of specific data sets and/or the
increased relevance of other theoretical corrections (such as power-suppressed
corrections), is left to future study.

The deterioration of the quality of the pion FF fit upon reduction of
the value of $Q_{\rm cut}$ has also partly been observed in
Ref.~\cite{Borsa:2022vvp}. There, however, the NNLO fit became worse than its
NLO counterpart only for values of the cut on the virtuality
$Q_{\rm cut}\lesssim\sqrt{2.00}$~GeV. This is different from what we observe.
Even if the analysis in Ref.~\cite{Borsa:2022vvp} and ours are based on a
similar data set, they however differ for the FF parametrisation and
optimisation methodology. Understanding the origin of the discrepancy between
the two sets of results would require a careful benchmark which goes beyond the
scope of this paper.

However, the results in Ref.~\cite{Borsa:2022vvp} and ours question
whether, at such small values of $Q$, the leading-power factorisation framework
used to describe SIDIS measurements is reliable. Poorly known power
corrections, including interplay between initial- and final-state
hadron mass effects~\cite{Accardi:2014qda,Guerrero:2015wha}, may become
dominant, or the fragmentation regime may not even
hold~\cite{Boglione:2016bph,Boglione:2019nwk,Boglione:2022gpv}.
For all of these reasons, while the fit quality --- as quantified by the value
of the $\chi^2$ per data point --- may remain acceptable for values of
$Q_{\rm cut}$ smaller than the one chosen as default in our fits
($Q_{\rm cut}=2$~GeV), we consider that the latter remains 
conservative against these effects. We have explicitly checked that the lower
the value of $Q_{\rm cut}$, the larger the distortion of the FFs (up to a couple
of standard deviations at intermediate values of $z$ for quark FFs), for either
pions or kaons and irrespective of the perturbative order, in comparison to
those displayed in Figs.~\ref{fig:pions_NLOvsNNLO}
and~\ref{fig:kaons_NLOvsNNLO}.

The second study consists in investigating the role of the correlation model
adopted to analyse the COMPASS SIDIS data. As already mentioned, in contrast
to our previous analysis~\cite{Khalek:2021gxf}, we no longer assume the entirety
of systematic uncertainties to be 100\% correlated, but only a fraction of them
equal to 80\%. The remaining fraction of each systematic uncertainty is treated
as fully uncorrelated, and is added in quadrature to the statistical
uncertainty. This correlation model was singled out in the papers in which the
COMPASS measurements were presented~\cite{COMPASS:2016crr,COMPASS:2016xvm}. As
already mentioned, the effect of this change is a significant reduction of the
$\chi^2$ per data point in comparison to the NLO pion fit of
Ref.~\cite{Khalek:2021gxf}; FFs are affected by fluctuations not large than
a standard deviation, as one can infer by comparing
Fig.~\ref{fig:pions_NLOvsNNLO} with Fig.~6 in Ref.~\cite{Khalek:2021gxf},
and as we have explicitly checked.

That being said, we repeated our NLO and NNLO fits for the pion FFs with two
alternative decorrelation models: one in which the systematic
uncertainties of the COMPASS data are 100\% correlated; and one in which the
systematic uncertainties in the COMPASS data are 100\%
uncorrelated. The fits are repeated for each value of $Q_{\rm cut}$ considered
above. Our aim is to investigate whether the NLO and NNLO values of the
global $\chi^2$ per data point follow the same pattern observed in our default
fits as $Q_{\rm cut}$ is varied.

The results are displayed in the left panel of Fig.~\ref{fig:chi2_scan}, from
which we draw two observations. First, the fit quality of the NNLO fit always
remains worse than that of the NLO fit, irrespective of the correlation model
used in the COMPASS data and of the value of $Q_{\rm cut}$. Second, the
correlation model affects the fit quality significantly: as the amount of
correlation increases, not only the value of the global $\chi^2$ per data point
becomes higher, but also the deterioration of the fit quality
occurs at larger values of $Q_{\rm cut}$. Furthermore, we have explicitly checked
the effect of the decorrelation model on the fitted FFs. We have generally
found that, irrespective of the value of $Q_{\rm cut}$ and of the perturbative
order, pion FFs vary very little in comparison to our default if the systematic
uncertainties in the COMPASS measurements are treated as fully uncorrelated.
In particular, the variation is significantly smaller than that due to NNLO
corrections, see Fig.~\ref{fig:pions_NLOvsNNLO}. Note however that differences
in the $\chi^2$ per data point may be similar. For instance, with
$Q_{\rm cut}=2$~GeV, the difference in $\chi^2$ per data point between the
default NLO and NNLO fits is 0.08; the same difference between the NNLO
fits, in which the COMPASS systematic uncertainties are either partly
correlated (our default) or fully uncorrelated, is 0.09.
Distortions appear if the same systematic
uncertainties are treated as fully correlated. However, the size of the
distortion between fits with the same value of $Q_{\rm cut}$ typically does
not exceed one standard deviation. These results illustrate the paramount
importance of a careful estimate of experimental correlations --- and of their
proper treatment in the fit --- to correctly interpret the fit quality in
terms of $\chi^2$ per data point.

\section{Summary and outlook}
\label{sec:conclusions}

In this paper we have extended the determination of NLO
pion FFs of Ref.~\cite{Khalek:2021gxf} in two respects. First, pion
SIA and SIDIS measurements have now been analysed up to NNLO accuracy in
perturbative QCD. Second, we have also determined companion kaon FF sets.
Our study is based on a consolidated framework that combines a neural-network
parametrisation of FFs with a Monte Carlo representation of their
uncertainties. This framework ensures that model bias is reduced as much as
possible, and that experimental and PDF uncertainties are faithfully
propagated into FFs.

We have found that inclusion of NNLO corrections does not improve the
quality of the pion FF fit, as measured by the $\chi^2$ per data point, but
it does for the kaon FF fit. Although the modifications of the NNLO FFs are
qualitatively as expected with respect to the NLO FFs, the reason for the
quantitative behaviour requires further investigations, possibly in the context
of a benchmark with other FF sets, such as those determined
in Ref.~\cite{Borsa:2022vvp}. As also noted in Ref.~\cite{Borsa:2022vvp},
poorly known power corrections, beyond the leading-twist factorisation
formalism used here, may play a role in the kinematic region covered by current
SIDIS measurements. Indeed, we have observed a fast deterioration of the fit
quality if the cut on the virtuality of the SIDIS process is lowered,
with the deterioration in the NNLO fit being more remarkable than in the NLO
fit. We have finally exposed the importance of a correct estimation and
treatment of experimental correlations to interpret the fit quality in terms of
the $\chi^2$ per data point.

Our NNLO pion and kaon FF sets, being the only ones to be publicly delivered to
date, could be used in a number of computations that require a matching
perturbative accuracy. For example, to make predictions of SIDIS cross sections
measured by the future EIC at higher energy. Or to determine, for the first
time at NNLO, longitudinally polarised PDFs from a simultaneous analysis of
polarised inclusive deep-inelastic scattering and SIDIS measurements. Or else
to serve as baseline for the parametrisation of transverse-momentum-dependent 
FFs.

\vspace{1cm}
\noindent The results presented in this paper have been obtained with the
public code available in Ref.~\cite{rabah_abdul_khalek_2022_6264693}, see
\begin{center}
\href{https://github.com/MapCollaboration/MontBlanc}{https://github.com/MapCollaboration/MontBlanc}.
\end{center}
For each perturbative order and hadron species (pion and kaon), we deliver the
FF sets for the positively charged hadrons, for the negatively charged hadrons
and for their sum. The names of the FF sets are as follows:
\begin{itemize}
\item NLO, pion:
  {\tt MAPFF10NLOPIp}, {\tt MAPFF10NLOPIm}, {\tt MAPFF10NLOPIsum};
\item NNLO, pion:
  {\tt MAPFF10NNLOPIp}, {\tt MAPFF10NNLOPIm}, {\tt MAPFF10NNLOPIsum};
\item NLO, kaon:
  {\tt MAPFF10NLOKAp}, {\tt MAPFF10NLOKAm}, {\tt MAPFF10NLOKAsum};
\item NNLO, kaon:
  {\tt MAPFF10NNLOKAp}, {\tt MAPFF10NNLOKAm}, {\tt MAPFF10NNLOKAsum}.
\end{itemize}
These FF sets are available from Ref.~\cite{rabah_abdul_khalek_2022_6264693},
where notebooks containing reports of the fits are also provided, and from the
{\sc LHAPDF} library~\cite{Buckley:2014ana}. Note that the NLO pion FF sets
replace those delivered in the previous paper~\cite{Khalek:2021gxf}.

\input{acknowledgements.tex}

\bibliography{MAPFFNNLO}

\end{document}

%% file: _tables/chi2.tex
\begin{tabularx}{\textwidth}{|Xc|rcc|rcc|}
  \toprule
  &
  & \multicolumn{3}{c|}{$h=\pi$}
  & \multicolumn{3}{c|}{$h=K$} \\
    \multirow{2}{*}{Experiment}
  & \multirow{2}{*}{ Ref.}
  & \multirow{2}{*}{$N_{\rm dat}$}
  & $\chi^2/N_{\rm dat}$
  & $\chi^2/N_{\rm dat}$
  & \multirow{2}{*}{$N_{\rm dat}$}
  & $\chi^2/N_{\rm dat}$
  & $\chi^2/N_{\rm dat}$
  \\
  & & & NLO & NNLO & & NLO & NNLO \\
  \midrule
  BELLE $h^\pm$                  & \cite{Belle:2013lfg} &
   70 & 0.14 & 0.13 &  70 & 0.39 & 0.41 \\
  BABAR $h^\pm$                  & \cite{BaBar:2013yrg} &
   39 & 0.91 & 0.76 &  28 & 0.36 & 0.25 \\
  TASSO 12 GeV $h^\pm$           & \cite{TASSO:1980dyh} &
    4 & 0.90 & 0.92 &   3 & 0.85 & 0.87 \\
  TASSO 14 GeV $h^\pm$           & \cite{TASSO:1982bkc} &
    9 & 1.33 & 1.35 &   9 & 1.24 & 1.22 \\
  TASSO 22 GeV $h^\pm$           & \cite{TASSO:1982bkc} & 
    8 & 1.65 & 1.81 &   6 & 0.89 & 0.90 \\
  TPC $h^\pm$                    & \cite{TPCTwoGamma:1988yjh} & 
   13 & 0.23 & 0.25 &  13 & 0.38 & 0.40 \\
  TASSO 30 GeV $h^\pm$           & \cite{TASSO:1980dyh} &
    2 & 0.30 & 0.34 & --- & ---  & ---  \\
  TASSO 34 GeV $h^\pm$           & \cite{TASSO:1988jma} &
    9 & 1.08 & 1.48 &   5 & 0.07 & 0.06 \\
  TASSO 44 GeV $h^\pm$           & \cite{TASSO:1988jma} &
    6 & 1.13 & 1.37 & --- & ---  & ---  \\
  TOPAZ $h^\pm$                  & \cite{TOPAZ:1994voc} &
    5 & 0.24 & 0.37 &   3 & 0.10 & 0.11 \\
  ALEPH  $h^\pm$                 & \cite{ALEPH:1994cbg} &
   23 & 1.24 & 1.46 &  18 & 0.49 & 0.48 \\
  DELPHI (inclusive) $h^\pm$     & \cite{DELPHI:1998cgx} & 
   21 & 1.31 & 1.25 &  23 & 0.97 & 0.99 \\
  DELPHI ($uds$ tagged) $h^\pm$  & \cite{DELPHI:1998cgx} &
   21 & 2.68 & 2.89 &  23 & 0.44 & 0.38 \\
  DELPHI ($b$ tagged) $h^\pm$    & \cite{DELPHI:1998cgx} &
   21 & 1.58 & 1.73 &  23 & 0.42 & 0.45 \\
  OPAL  $h^\pm$                  & \cite{OPAL:1994zan} &
   24 & 1.63 & 1.79 &  10 & 0.39 & 0.36 \\
  SLD (inclusive) $h^\pm$        & \cite{SLD:2003ogn} &
   34 & 1.05 & 1.13 &  35 & 0.83 & 0.67 \\
  SLD ($uds$ tagged) $h^\pm$     & \cite{SLD:2003ogn} &
   34 & 1.59 & 2.16 &  35 & 1.37 & 1.52 \\
  SLD ($b$ tagged) $h^\pm$       & \cite{SLD:2003ogn} &
   34 & 0.55 & 0.68 &  35 & 0.75 & 0.77 \\
  \midrule
  Total SIA              &   &
  377 & 1.03 & 1.15 & 339 & 0.58 & 0.57 \\
  \midrule
  HERMES $h^-$ $d$               & \cite{HERMES:2012uyd} &
    2 & 0.41 & 0.32 &   2 & 0.18 & 0.13 \\
  HERMES $h^+$ $p$               & \cite{HERMES:2012uyd} &
    2 & 0.01 & 0.02 &   2 & 0.05 & 0.04 \\
  HERMES $h^-$ $d$               & \cite{HERMES:2012uyd} &
    2 & 0.17 & 0.11 &   2 & 0.58 & 0.48 \\
  HERMES $h^+$ $p$               & \cite{HERMES:2012uyd} &
    2 & 0.35 & 0.32 &   2 & 0.56 & 0.43 \\
  COMPASS $h^-$                  & \cite{COMPASS:2016crr,COMPASS:2016xvm} &
  157 & 0.48 & 0.55 & 156 & 0.74 & 0.59 \\
  COMPASS $h^+$                  & \cite{COMPASS:2016crr,COMPASS:2016xvm} &
  157 & 0.62 & 0.72 & 156 & 0.76 & 0.67 \\
  \midrule
  Total SIDIS & & 
  322 & 0.47 & 0.52 & 320 & 0.64 & 0.54 \\
  \midrule
  {\bf Global data set}  & & {\bf 699} & {\bf 0.68} & {\bf 0.76}
                           & {\bf 659} & {\bf 0.62} & {\bf 0.55} \\
  \bottomrule
\end{tabularx}

%% file: acknowledgements.tex
\section*{Acknowledgments}
We thank Werner Vogelsang for discussions on the NNLO corrections to SIDIS.
V.~B. is supported by the European Union's Horizon
2020 research and innovation programme under grant agreement
\textnumero~824093. E.~R.~N. is supported by the UK STFC grant
ST/T000600/1. This material is based upon work supported by the U.S. Department
of Energy, Office of Science, Office of Nuclear Physics under contract DE-AC05-06OR23177. 
R. A. K. was partially supported by Nobuo Sato's grant supported by the DOE, 
Office of Science, Office of Nuclear Physics in the Early Career Program.